# The Politics Attention Makes: Platform Media Logic and the Mediatization of Politics

Petter Törnberg

Institute for Logic, Language and Computation (ILLC), University of Amsterdam. p.tornberg@uva.nl

**Abstract**

Empirical research on social media and politics has primarily treated platforms as distributive systems that expose users to particular messages. The mediatization literature, however, suggests shifting attention upstream: from circulation to production. Under intense competition for platform attention, political actors who depend on visibility face pressure to learn from recurrent differences in reach and engagement – shaping politics around platform media logic. This paper examines that production-side dimension of platforms political impact by introducing *attention price analysis*: an exploratory method for estimating the differentiated attention returns associated with forms of expression. Using RoBERTa reward models trained on residualized engagement across X/Twitter, Bluesky, and Mastodon, the analysis compares how platform environments reward rhetorical, emotional, epistemic, and relational features of public communication. The attention signal differs sharply across platforms and engagement actions. X/Twitter sharing rewards antagonism while penalizing respect and nuance; Bluesky reposting favors neutral, lower-emotion language; and Mastodon boosts reward reasoning, nuance, compassion, and collective expression. Toxicity is rewarded across platforms, but in bounded and nonlinear ways. The findings suggest that moving from X/Twitter to less engagement-optimized alternatives such as Bluesky and Mastodon does not eliminate attention pressures, but it may reward less antagonistic and more deliberative forms of politics. The paper contributes a production-side approach to social media and politics by making one dimension of platform media logic empirically visible.

***Keywords**:* attention economy; political communication; platformization; mediatization; engagement; amplification; reward models

## Introduction

The dominant empirical account of social media and democracy is a story about distribution. Algorithms rank, feeds curate, networks cascade, and users encounter some messages rather than others. Platforms are seen as machines — or "prisms" (Bail, 2022) — for selecting what reaches audiences, and the central political question is what this exposure does to citizens: whether it polarizes, misinforms, mobilizes, or demobilizes them (Bakshy et al., 2015; Lorenz-Spreen et al., 2023; Tucker et al., 2018). The focus on audience means that the democratic crisis associated with social media tends to be studied as primarily a crisis of the electorate (Törnberg, 2026a). This literature is large, technically sophisticated, and substantively important. But its findings are often less straightforward than early diagnoses implied: large-scale interventions that changed social media use, feed ranking, or partisan exposure altered what people encountered but often produced surprisingly limited effects on political attitudes, affective polarization, or offline political behavior (Allcott et al., 2020; Guess et al., 2023; Nyhan et al., 2023). Algorithms clearly shape what people see, but they do not straightforwardly determine what people believe, feel, or how they vote.

A long tradition in media and communication theory suggests that this distributional focus misses a key relationship between media and politics. Media do not merely transmit political messages; they reorganize the incentives and conditions under which messages are made, circulated, and rendered consequential (Couldry & Hepp, 2017; McLuhan, 1964; Postman, 1985). The *mediatization* literature has developed this argument most directly, showing how political actors, institutions, and movements adapt to media formats, rhythms, and evaluative criteria, until media logic becomes partly constitutive of how politics is organized and performed (Hjarvard, 2008; Strömbäck, 2008). Politics is not made elsewhere and then filtered through media, but it is produced in a field already structured by media: by what can become visible and command attention, and hence what is politically possible.

While media attention has always been a vital resource for politics, platforms have transformed attention from an amorphous and elusive condition of public relevance into a tangible, quantified, continuously updated signal. Likes, reposts, replies, views, and impressions make attention visible at the level of individual messages. They make it comparable across messages, accumulable across time, and convertible into influence, status, organizational resources, and sometimes income (Franck, 2019; Hwang, 2020; Pedersen et al., 2021). Platforms thereby help constitute attention markets: environments in which political expressions receive different expected returns, and in which actors who depend on visibility face pressure to learn from those returns. In this perspective, attention returns are sociotechnical. They emerge through the interaction of interface affordances, metric visibility, recommender systems, user cultures, audience attention heuristics, community norms, actor reputation, and external events. Mediatization hence acts through a composite attention environment rather than through a single platform mechanism.

While the mediatization literature has emphasized production-side impacts of media on politics, the empirical social media literature has tended to focus on distribution and exposure. The reason is in part methodological: mediatization describes large-scale structural changes that are challenging to operationalize. The empirical social media literature often observes what spreads, but not what producers are rewarded for making—and treats engagement as a single undifferentiated signal. Existing research identifies particular features associated with diffusion, or studies engagement as an outcome, but it rarely estimates the broader reward structure confronting political producers.

To address this gap, this paper introduces *attention price analysis*: an exploratory method for estimating the attention-price signals that platformed communication environments present to content producers. It uses expected market returns as an analytic metaphor to offer a production-side complement to exposure-centered research and brings the mediatization thesis into the empirical study of social media by making one part of media logic observable as differentiated attention returns. If platforms organize public communication as attention markets, then content features should carry systematically different expected returns across engagement outcomes and across platforms with different architectures. The key object is the *attention price function*: the expected return, in engagement and reach, to different features of expression that is defined by the platforms' assemblage of algorithms, audiences, cultures, reputation, and affordances.

Operationally, we fine-tune RoBERTa reward models on residualized engagement — net of audience size and timing — across three contrasting platform environments: X/Twitter, Bluesky, and Mastodon. We then interpret the resulting reward functions through theoretically motivated textual features and use a rewriting experiment to probe what happens to political content when optimized under different platform-specific price functions. The rewriting experiment is an exploratory mechanism probe, asking which directions of transformations become rewarded when predicted attention becomes the criterion of revision.

The analysis yields five findings. First, text alone carries a detectable price signal: it predicts residualized engagement across X/Twitter, Bluesky, and Mastodon. Second, approval and circulation are distinct markets. Approval, measured through likes, is comparatively portable across platforms and rewards personal, positive, and affiliative expression. Third, circulation is more platform-specific. URL-free X/Twitter sharing shows the clearest antagonistic grammar, rewarding insult, identity attack, and toxicity while penalizing respect, positive emotion, reasoning, nuance, and personal narrative. Bluesky reposts instead favor a more neutral, lower-emotion register, while Mastodon boosts reward reasoning, nuance, compassion, and collective expression, with smaller contentious returns. X/Twitter impressions form a separate reach profile, favoring more analytical language rather than overt antagonism. Fourth, the computational rewriting probe suggests that semantic fidelity constrains attainable predicted reward within the learned reward functions: attention-optimized candidates gain substantially more predicted reward than fact-preserving candidates but drift toward personalization, moral-emotional framing, social positioning, and incivility. Fifth, Mastodon's strong text-attention gradient indicates that attention pressure is not reducible to centralized recommender algorithms, although Mastodon and Bluesky reward less antagonistic and more deliberative forms of circulation than X/Twitter in important respects.

These findings document an observed price structure that assigns different returns to rhetorical forms associated with contemporary political conflict: identity-marking, emotional charge, antagonism, and qualification. Attention price analysis does not identify the causal consequences of social media for political production. It offers the first steps for making the incentive environment empirically visible: what kinds of expression platformed attention makes advantageous, what forms it makes costly, and how those returns differ across approval, circulation, and reach.

## From Media Logic to Attention Prices

The computational and empirical literature on social media and politics has been especially effective at studying circulation. It measures what users encounter, how content diffuses, how networks sort publics, how recommender systems rank information, and how exposure relates to polarization, misinformation, participation, and trust. The same orientation underlies concerns with echo chambers and filter bubbles, where the central worry is that networks, personalization, and algorithmic curation narrow the range of political information citizens encounter. In this platform-effects tradition, social media's political significance is often operationalized through distribution: platforms matter because they shape the flow of information to citizens. Recommender systems optimize for engagement; engagement is associated with outrage, moralization, and partisan hostility; divisive content may therefore travel farther and expose users to more polarizing political material (Brady et al., 2017; Rathje et al., 2021). On this account, platforms are dangerous because they are biased distributors, and the democratic crisis is cast primarily as a crisis of the voting public: citizens see the wrong content, become more hostile, and carry those dispositions into democratic life (Mason, 2018).

This distributional paradigm has generated a large and sophisticated body of research, but its causal implications remain contested. Large-scale experiments that altered platform exposure or temporarily removed social media altogether have not produced the expected reductions in polarization (Allcott et al., 2020). The Meta studies from the 2020 U.S. election, for instance, found that disabling algorithmic ranking or adjusting exposure to partisan sources changed what users saw on Facebook and Instagram in measurable ways, but produced little detectable change in political attitudes, affective polarization, or offline political behavior (Guess et al., 2023; Nyhan et al., 2023).

## Mediatization of Politics

A longer tradition in media and communication theory, however, suggests a fundamentally different understanding of how social media impact political life. Media do not only distribute political communication; they reshape the conditions under which political communication is produced (Hjarvard, 2008; Strömbäck, 2008). The mediatization tradition begins from this proposition: political actors speak, organize, campaign, and perform in anticipation of media visibility. Broadcast media rewarded the sound bite, the staged event, the charismatic leader, and the televisual performance – in short, politics as entertainment (Postman, 1985). Political actors adapted not because television commanded them to do so, but because visibility became a condition of political relevance (Altheide & Snow, 1979; Chadwick, 2017). The conditions for visibility were defined by the medium, and media logic therefore entered political practice through candidate selection, campaign organization, institutional routines, and styles of public performance.

This is not merely a question of rhetoric or style. Media reshape political supply itself: which actors become viable, which issues are worth emphasizing, which conflicts become useful, and which forms of political identity can be mobilized effectively. Communication is not external to politics, since political styles, coalitions, identities, leadership forms, and repertoires of conflict are partly constituted through public performance.

Deep mediatization extends the mediatization argument beyond the adaptation of particular institutions to particular media formats. It describes the saturation of social life by interconnected, datafied media infrastructures, such that mediated forms become constitutive of how social reality is organized, perceived, and acted upon (Couldry & Hepp, 2017; Hepp, 2020). Under deep mediatization, there is no clearly separable "pre-media" politics. Political actors, journalists, movements, institutions, and publics operate inside infrastructures that shape what becomes visible, measurable, comparable, and actionable. The platformization literature specifies the institutional form this takes today: platforms do not simply relay existing cultural and political production, but reorganize it around programmable infrastructures, metrics, governance rules, affordances, and business models (van Dijck & Poell, 2013; Nieborg & Poell, 2018; Poell et al., 2019).

## Attention Markets

Social media intensify mediatization by constituting what has been referred to as an 'attention market' (Pedersen et al., 2021; Törnberg 2025). Politics has always been shaped by struggles for attention, but platforms transform attention from an elusive and abstract sense of public relevance into a quantified, continuously updated, and publicly visible currency. Likes, shares, reposts, replies, views, and impressions make audience response observable at the level of individual messages. What expressions are rewarded on a particular platform is an expression of an inextricable assemblage of recommender ranking, affordances, algorithms, audience, culture, actor status, and exogenous factor. At the same time, platforms make attention comparable across messages, accumulable across time, and convertible into influence, status, organizational resources, and even income (Törnberg 2025). Franck (2019) theorizes this as the emergence of attention as a parallel currency system: attention becomes a scarce and valuable resource with its own exchange logic. Hwang (2020) pushes this further by showing how digital infrastructures turn attention into a market object, governed by mechanisms of valuation, mispricing, speculation, and instability. Applied to politics, platforms constitute markets in which political expression is differentially valued. In the terms of mediatization theory, these valuations are one observable component of platform media logic: they indicate which communicative forms become more or less advantageous within a given media environment.

This provides a mechanism through which media logic enters the production of politics. Mediatization theory argues that political actors adapt to the formats, rhythms, and evaluative criteria of media systems (Altheide & Snow, 1979; Hjarvard, 2008; Strömbäck, 2008). Attention markets specify how this adaptation occurs under platform conditions. Through social media, political actors, journalists, movements, and institutions encounter recurring signals about which expressions travel, which tones provoke response, which identities mobilize publics, and which forms of conflict command visibility.

Beyond their intrinsic political value, attention metrics also operate as representational signals. Likes, reposts, replies, and views are often read as evidence of what “people” think, feel, and value. In this sense, platforms provide a distorted social mirror through which producers, journalists, political actors, and ordinary users infer public opinion, issue importance, and the social legitimacy of particular positions. This links attention markets to older concerns with perceived public opinion, impersonal influence, and the spiral of silence (Noelle-Neumann, 1974; Mutz, 1998; Gunther, 1998), as well as newer work on social media metrics as cues of public sentiment and social endorsement (Messing & Westwood, 2014; McGregor, 2019). Yet the public made visible through platforms is shaped by unequal participation, strategic amplification, platform-specific audiences, and engagement-biased visibility (Mellon & Prosser, 2017). Attention markets can therefore create a double distortion: they reward some forms of political expression more than others, and then present the resulting pattern of visibility as if it were a reflection of what the public actually thinks.

By making engagement metrics visible and consequential, platforms help constitute attention markets in which political expressions acquire differential returns (Pedersen et al., 2021). The logic resembles what Marx (1867/1976) identified as the *disciplinary force of market competition*: markets need not command producers directly, but create selection environments in which actors who fail to respond to price signals risk being outcompeted, marginalized, or displaced. Platformed attention markets operate in a similar way. Political actors, journalists, movements, and institutions do not need to consciously maximize engagement, or understand the technical operation of a platform, to learn which styles appear to work. They encounter the pressure through recurring differences in reach, engagement, public reaction, professional recognition, and organizational relevance. The result is not merely a change in political messaging, but a selection environment for politics itself: one that helps determine which actors gain visibility, which issues and conflicts become useful, which identities become performable, and which forms of politics can thrive.

The way content producers are adapting to this signal can be seen in qualitative studies of digital production. Cotter (2019) describes creators as engaged in a continuous "visibility game," inferring platform rules from fluctuating metrics and adapting through trial, error, and imitation. Mears (2023) shows that social media production is organized through a broader cultural economy of attention in which producers learn practical techniques and evaluative orientations. In journalism, Christin (2020) and Petre (2021) document how analytics dashboards recast editorial judgment in the language of measurable audience response, reshaping how news workers understand value, success, and professional performance. These studies show how producers respond to the implicit reward signal produced by platforms, becoming part of the environment within which actors learn what counts, what travels, and what appears worth doing.

The mediatization perspective implies a shift in our understanding of our current democratic crisis. Whereas distributional accounts cast the democratic crisis associated with social media as primarily a crisis of the electorate, mediatization shifts the focus to political elites (Törnberg, 2026a). It highlights not primarily voter polarization, but elite misinformation, democratic backsliding, and forms of politics that divide society into opposed camps, attach moral character to group identity, and transform disagreement into evidence of threat, corruption, or betrayal (Finkel et al., 2020; Iyengar et al., 2019; Mudde & Kaltwasser, 2017; Törnberg & Chueri 2025). The argument is not that platforms invented these forms of politics, but that they contribute to an attention environment within which they are awarded. If antagonistic identity claims repeatedly earn circulation, they become more advantageous to produce and more available for imitation. If moral certainty travels better than qualification, communicators face a recurring tension between precision and visibility. If hostile performance generates more amplification than institutional compromise, actors who specialize in conflict may gain a structural advantage in the struggle for public relevance. In this way, attention markets may help select not only messages, but political movements and actors.

## Measuring the Attention Gradient

One reason for the lack of empirical research oriented around mediatization is the challenges involved in its operationalization. The claim that politics adapts to media logic is difficult to capture empirically, as media logic operates as a diffuse institutional force: a set of formats, rhythms, norms, and evaluative criteria that shape political conduct over time. The attention market metaphor however also suggests an opening for the empirical exploration of mediatization. If platformed visibility is structured through recurring returns to expression, then media logic can be studied as a reward structure: the expected attention return attached to particular rhetorical forms, emotional registers, identity claims, and modes of conflict.

Drawing from the distribution literature, there are already clues of what this reward structure privileges. Moral-emotional language increases the diffusion of political content, with Brady et al. (2017) estimating that each additional moral-emotional word increases sharing by roughly twenty percent. Rathje et al. (2021) show that references to political out-groups are especially strong predictors of sharing, with out-group language producing substantially more diffusion than in-group language. Broader reviews of social media and democracy likewise identify recurring links among engagement, affective polarization, partisan hostility, and divisive political content (Lorenz-Spreen et al., 2023). These findings are often read through a distributional lens: they tell us which content travels and what users may encounter, but they also imply a production-side gradient, as communicators observing these returns receive evidence about what is advantageous to create.

Existing research, however, rarely seeks to map the attention market confronting political producers. Platform-effects research examines exposure and downstream attitudes, but not the incentives producers face before content enters circulation. Diffusion studies identify features associated with sharing but rarely reconstruct the broader reward structure across platforms and engagement outcomes. Ethnographic work documents adaptation to metrics but usually cannot quantify the systematic differences in relative returns to different rhetorical forms. What is missing is a way to translate the mediatization claim into an estimable object: a measure of how platform environments differentially value political expression.

We call the expected return, in engagement or visibility, to a given form of expression the *attention price function*. The concept translates the media-theoretic idea of platform logic into a measurable object, while treating that logic as sociotechnical – arising from a combination of recommender ranking, algorithms, affordances, culture, and audiences. The price function captures what different expressive features are worth in the observed attention market, how their value differs between approval and amplification, and how these values vary across platform environments. The analysis that follows cannot establish how social media have reshaped politics, nor can it isolate whether a given return arises from recommender ranking, affordances, audience culture, actor status, or exogenous events. It can establish whether the attention signal is structured, what forms of expression it rewards, and where platformed environments differ.

## Data and Methods

The three corpora were selected to contrast platformed communication environments. X/Twitter combines centralized engagement-optimized ranking and a contentious culture. The raw corpus contains 2.34 million English-language posts collected between October 2023 and September 2025 through repeated searches for common English stop words at randomly sampled moments. After removing pure retweets and applying text-quality filters, 812,979 posts remained for modeling.

Bluesky combines portable identity and user-selectable feeds with an open social graph. We use the high-coverage corpus described by Failla and Rossetti (2024) and released through Zenodo (Failla & Rossetti, 2025). The modeling sample contains 500,000 English-language posts, with follower counts reconstructed from the public follow graph. The underlying dataset covers posts through March 2024, so the platform observation windows are not contemporaneous.

Mastodon is decentralized across independently governed instances and relies heavily on chronological follower and local timelines rather than one centralized engagement-ranking system. Because an available archival corpus did not provide the post-level engagement measures required here, we collected public local-timeline posts from large English-language instances. The prepared table contains 950,050 posts with favourites, boosts, replies, follower counts, and timestamps. Account identifiers were HMAC-SHA256 hashed before storage; usernames, display names, avatars, and biographies were not retained.

All corpora contain general public communication rather than only posts classified as political. This choice follows from the paper's theoretical object: platform-specific visibility regimes, not the language of political actors alone. Political actors, journalists, activists, organizations, and ordinary users compete for attention within the same broader communicative environments, and the reward structures they encounter are therefore not confined to explicitly political posts. The analysis thus estimates the attention signals of the wider platform field into which political communication enters, and from which actors may infer platform-specific visibility norms. Posts shorter than 20 characters, pure reposts, and URL-only or mention-only posts were excluded. Additional collection and filtering details appear in the Supplemental Materials.

***Outcomes and Reward Models***

The models do not estimate the causal effect of isolated textual choices, nor do they identify the separate contribution of algorithms, affordances, audiences, and community norms. They estimate the composite attention signal confronting producers in the observed platform environment. We estimate three primary outcomes for each platform: likes, shares, and their sum. Shares comprise retweets and quote tweets on X/Twitter, reposts on Bluesky, and boosts on Mastodon. On X, we moreover estimate impressions. Within each platform and outcome, the log-transformed engagement measure is residualized on log follower count and indicators for hour, weekday, and calendar month. The residual is the reward-model target. This adjustment removes major audience and timing advantages but does not fully measure effective exposure. Follower counts are recorded at collection and cannot capture feed ranking, audience activity, or community embeddedness.

We fine-tune RoBERTa-base regressors (Liu et al., 2019) on post text alone. Models use a maximum sequence length of 192 tokens, AdamW with a learning rate of 2e-5, effective batch size 32, two epochs, weight decay .01, and a .06 warmup ratio. Deterministic post-identifier hashing produces approximately 80/10/10 training, validation, and test partitions. Pearson and Spearman correlations on untouched test sets measure predictive accuracy.

Two X/Twitter extensions separate active engagement from reach and rhetoric from link distribution. An impressions model predicts recorded post views. A URL-free shares model is retrained on posts without external links because linked news and institutional material may circulate for source value rather than wording. Removing URL-bearing posts reduces X/Twitter's modeling corpus by 26.0%, compared with 3.7% for Bluesky and 47.7% for Mastodon. Impressions are analyzed as a distinct reach outcome rather than assumed to share the rhetorical grammar of reposting.

Cross-platform transfer applies a model trained on one platform to held-out posts from another. Transfer efficiency is the transferred Pearson correlation divided by the target platform's own-platform correlation. Because target scales and populations differ, this analysis tests portability of rank-relevant textual signal rather than equality of engagement processes.

### *Interpretable Features*

We score up to 300,000 posts per platform with an interpretive feature battery chosen to cover dimensions that prior work connects to political attention and democratic communication: moral-emotional language and diffusion (Brady et al., 2017; Araque et al., 2020), out-group animosity and affective polarization (Rathje et al., 2021; Finkel et al., 2020; Iyengar et al., 2019), emotional expression (Demszky et al., 2020), and incivility, toxicity, and constructiveness in online discourse (Lees et al., 2022). The battery includes widely used surface features; VADER and NRC sentiment measures; MoralStrength; GoEmotions; Perspective API toxicity and constructiveness attributes; and social-cognitive markers. These measures are used as a structured vocabulary for interpreting reward models, not as definitive measurements of latent attitudes, motives, or deliberative quality. Both each feature and predicted reward are residualized on post length, URL and mention counts, hashtags, punctuation, capitalization, reply status, follower count, hour, weekday, and month.

Broad indices can conceal opposing associations among their components. We therefore audit all 85 measures individually and organize the interpretation around five theoretically motivated families: *relational expression* (personal story, affinity, respect, and compassion), *emotional register*, *epistemic style* (reasoning and nuance), *collective positioning*, and *incivility*. Figure 1 displays 13 nonredundant measures selected after the full audit to represent the largest theoretically distinct patterns; the complete coefficient matrix is reported in the Supplemental Materials. Because off-the-shelf classifiers can be sensitive to context, dialect, quotation, and topic, results are interpreted as convergent descriptive evidence across feature families rather than as validation of any single measure. Because incivility may have diminishing returns, toxicity is also analyzed by empirical decile and with controlled quadratic models.

### ***Computational Rewriting Probe***

The computational rewriting probe explores the directional pressure implied by the learned reward functions, and offer illustrative examples of the pressures of the reward functions. Nine constructed neutral statements cover mundane, news, and institutional language, and gpt-5-mini was employed to generate candidates in eight styles: sensational, angry, conflictual, populist, moralized, personal, positive, and informational. Each platform's combined-engagement model selects the candidate with the highest predicted reward.

Two regimes vary semantic constraint. Strict-factual candidates require sentence-embedding cosine similarity of at least .78 and content-word Jaccard overlap of at least .45. Attention-seeking candidates use thresholds of .65 and .30, permitting greater changes in emphasis and framing while retaining the topic. Sentence embeddings use all-MiniLM-L6-v2 (Reimers & Gurevych, 2019). The probe does not model human behavior, and candidates were not manually fact-checked. The aim is to illustrate what is selected under the estimated reward functions, not how actual political actors would rewrite messages or how audiences would respond in a controlled exposure experiment.

## Results

| Platform | Outcome | Pearson r | Spearman ρ |
|---|---|---|---|
| X/Twitter | Impressions | .524 | .496 |
| X/Twitter | Likes | .305 | .262 |
| X/Twitter | Shares | .364 | .199 |
| X/Twitter | Shares (no URL) | .241 | .159 |
| X/Twitter | Combined | .305 | .263 |
| Bluesky | Likes | .290 | .299 |
| Bluesky | Shares | .447 | .315 |
| Bluesky | Combined | .292 | .304 |
| Mastodon | Likes | .585 | .571 |
| Mastodon | Shares | .504 | .422 |
| Mastodon | Combined | .581 | .568 |

***Table 1****: Own-Platform Held-Out Reward-Model Accuracy. Combined engagement is likes + shares. The impressions model is available only for X/Twitter. The no-URL model was retrained on posts without external links.*

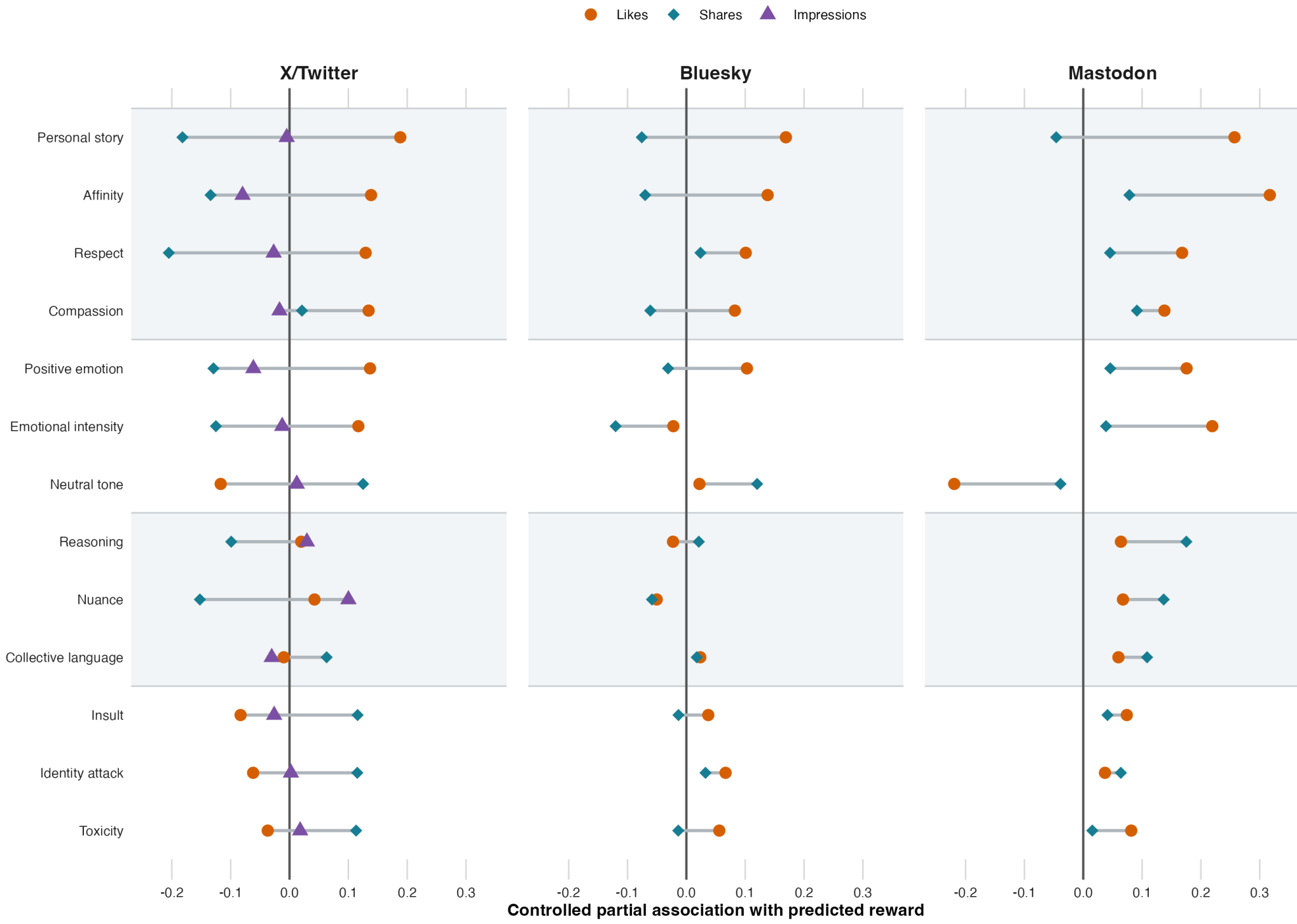

**Figure 1**: *A Shared Relational Approval Grammar, but Distinct Circulation and Reach Grammars. Controlled partial associations after surface, audience, and temporal controls. X/Twitter shares use the model retrained without external URLs; impressions are available only for X/Twitter. Bands group relational expression, emotional register, epistemic and collective style, and incivility. The full matrix appears in the Supplemental Materials.*

### *A Measurable, Partially Portable Price Signal*

Text predicts audience- and time-adjusted engagement on every platform (Table 1). Own-platform Pearson correlations range from .290 for Bluesky likes to .585 for Mastodon likes. On X/Twitter, impressions are predicted at r = .524, indicating that textual signal extends beyond active engagement to recorded reach. The URL-free X/Twitter shares model remains predictive (r = .241) but performs below the complete-corpus shares model (r = .364); linked-source distribution is therefore an important component of X/Twitter amplification, but not the entire signal. Impressions are not rhetorically interchangeable with shares: nuance (partial r = .100) and curiosity (.073) are positively associated with predicted views, whereas affinity (-.080) and positive emotion (-.061) are negative. Recorded reach therefore has a structured but more analytical profile than URL-free reposting.

Mastodon's combined reward is predicted at r = .581, despite the absence of a single centralized engagement-ranking feed. This suggests that strong text-attention coupling can emerge in a decentralized and largely chronological environment. Recommender systems may shape attention pressures, but centralized recommendation is not necessary for a measurable price signal.

Cross-platform transfer is positive but uneven. Off-diagonal Pearson transfer efficiencies range from approximately 5% to 61% of own-platform accuracy. The weakest transfer occurs when the Mastodon combined model is applied to X/Twitter; the strongest occurs when the Mastodon likes model is applied to Bluesky. Thus, the platforms share some textual reward structure, but no model provides a platform-independent attention function.

Feature-profile alignment clarifies where commonality lies. Like-reward profiles are strongly aligned between X/Twitter and Mastodon (r = .744) and moderately aligned for the other platform pairs (r = .376-.391). Share-reward profiles are less portable: X/Twitter and Mastodon are nearly uncorrelated (r = .011), while Bluesky and Mastodon are negatively aligned (r = -.101). Approval has a comparatively general social grammar; amplification depends more heavily on platform-specific meanings of recirculation.

### *Approval and Amplification Are Different Markets*

Figure 1 shows a shared relational approval grammar but sharply different circulation and reach profiles. Likes reward personal stories on X/Twitter (partial r = .188), Bluesky (.169), and Mastodon (.257), and affinity on all three (.139, .138, and .317, respectively). Respect, compassion, and positive emotion are also positively associated with likes across platforms. Emotional intensity is positively associated with likes on X/Twitter (.117) and especially Mastodon (.219), but is near zero on Bluesky (-.022). Approval is thus consistently relational, although its emotional register varies.

Amplification is more platform-specific. URL-free X/Twitter sharing displays the clearest antagonistic inversion: personal story (-.216), respect (-.210), affinity (-.167), reasoning (-.087), and nuance (-.145) are penalized, while insult, identity attack, and toxicity have partial associations between .107 and .123. Bluesky reposting instead favors a restrained register: neutral tone is positive (.120), emotional intensity is negative (-.120), and personal story, affinity, compassion, and nuance are modestly negative. Its linear incivility associations are small and mixed.

Mastodon boosting follows a third grammar. Reasoning (.175), nuance (.137), collective language (.108), compassion (.091), and affinity (.078) are positively associated with boosts. Personal story is slightly negative (-.046). Yet Mastodon is not purely pro-social: identity attack (.064) and insult (.041) also receive positive linear returns, and the sparse high-toxicity tail has elevated predicted reward. The dominant Mastodon pattern is reasoned, collective, and relational circulation with a smaller contentious component.

X/Twitter impressions form a fourth, distinct profile. Nuance (.100) and reasoning (.030) are positive, while affinity (-.080), positive emotion (-.061), and respect (-.027) are negative. Insult, identity attack, and toxicity are near zero or small. Recorded reach is therefore more analytical and less overtly antagonistic than URL-free sharing.

The full feature audit also qualifies the language of moralization. Moral-foundation lexicon measures are not a dominant or consistently positive observational predictor family. Moral-emotional change appears more clearly in the computational rewriting probe than as a common linear feature of naturally occurring high-reward posts.

The complete X/Twitter shares corpus also contains a broadcast-distribution economy: URLs, length, and institutional packaging strongly predict shares. The 85-feature interpretive battery explains 3.6% of predicted shares in the complete evaluation corpus. For the retrained URL-free shares model, it explains 11.1%. This result does not mean rhetoric replaces source value. It shows that once external links are removed, the remaining amplification signal is more legible as a rhetorical price function.

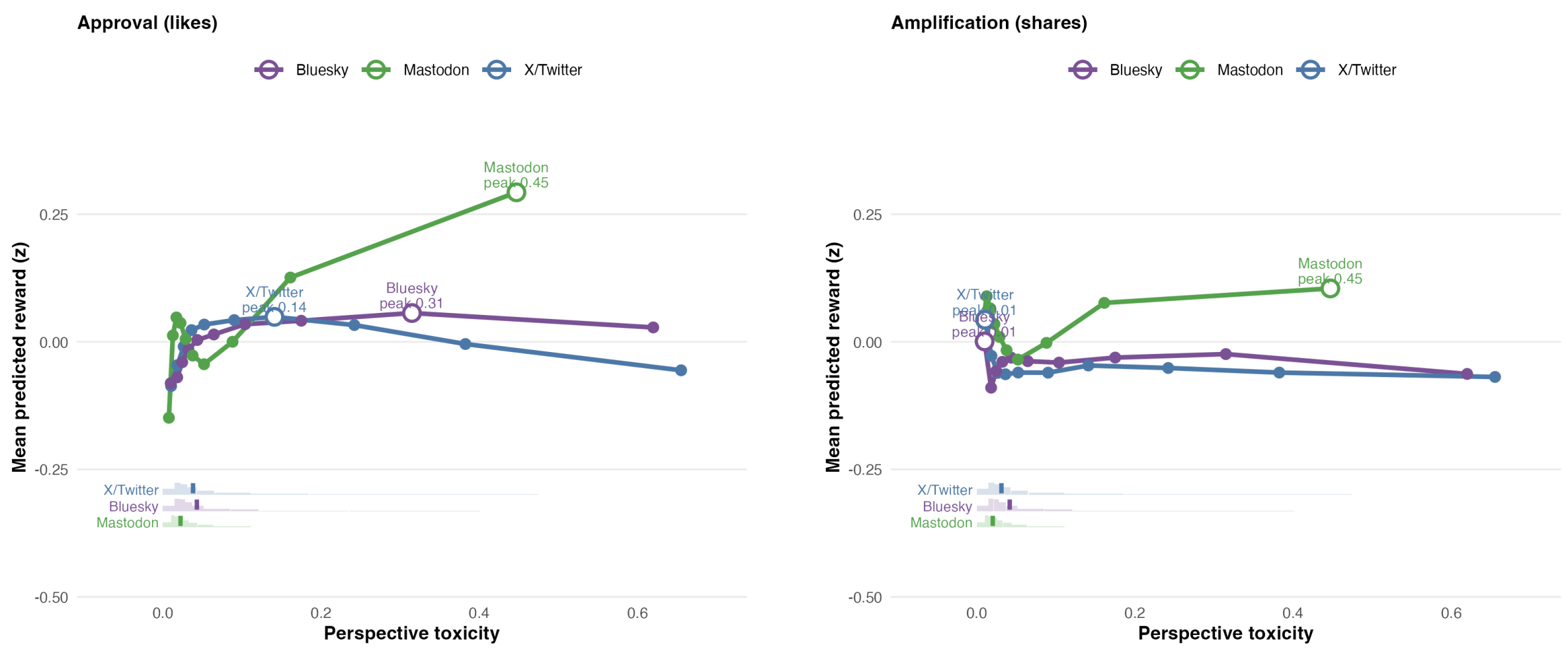

***Figure 2****: Incivility Has Nonlinear, Platform-Specific Returns. Mean predicted likes and shares reward by empirical Perspective toxicity decile. Open circles mark the observed peak; short ticks mark median toxicity among the top 1% of predicted-reward posts. Shaded bands show the toxicity distribution.*

### *Incivility Has Nonlinear Returns*

Linear averages however poorly represent the return on incivility. On X/Twitter, predicted combined and like reward rise through the sixth toxicity decile, where mean Perspective toxicity is approximately .14, and then decline. Bluesky combined and like reward peak later, near mean toxicity .31. These values are above the mass of routine platform discourse but below the most toxic tail (Figure 2).

The highest-reward posts are not the most abusive: X/Twitter's top 1% of combined-reward posts have median toxicity .039, compared with .065 in the full scored corpus. The evidence therefore supports a bounded antagonism premium, rather than that more toxicity is always better. Moderate antagonism can differentiate a message, while extreme language may encounter audience sanction, moderation, or narrower topical contexts.

Mastodon however differs. Its highest observed toxicity decile has elevated predicted reward for combined engagement and likes, producing an apparently monotonic upper tail. Highly toxic Mastodon posts are sparse, however, and the top-reward distribution remains concentrated at low toxicity. The pattern may reflect rare controversy or quoted political speech rather than a general optimum at severe toxicity.

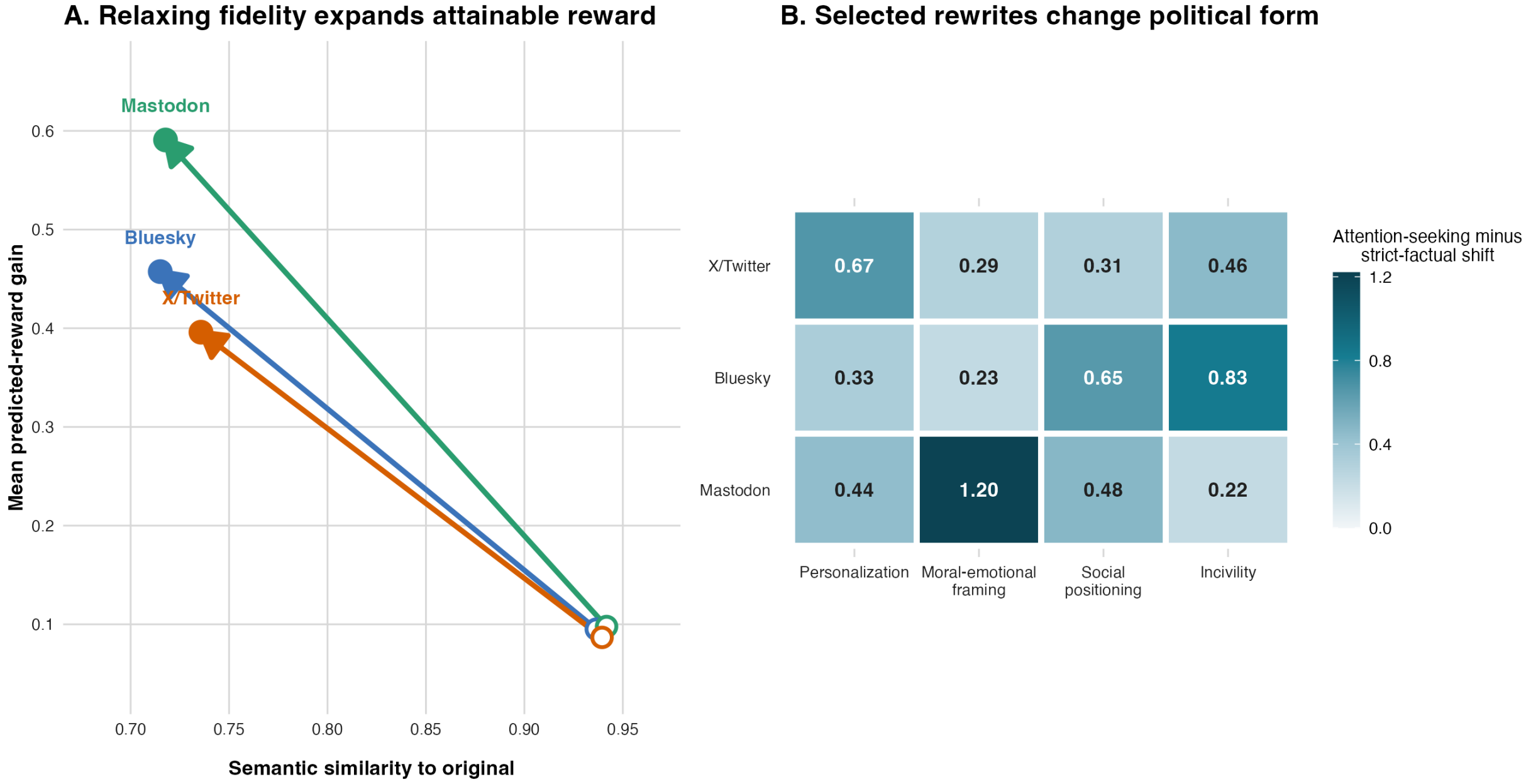

***Figure 3**: Attention Optimization Trades Fidelity for Reward and Changes Message Form. Panel A connects each platform's strict-factual mean to its attention-seeking mean; filled endpoints mark attention-seeking selection. Panel B reports standardized attention-seeking minus strict-factual shifts in rewrite-style indicators. These indicators describe generated-candidate changes and are not identical to the observational feature families in Figure 1.*

| Platform | Original | Selected attention-seeking output |
|---|---|---|
| X/Twitter | Participants may submit written comments during the consultation period. | I'm officially submitting a written comment during the consultation period—they can't ignore us if enough voices show up. |
| Bluesky | Researchers reported that the program increased participation among first-time applicants. | Enough of the same old setup—this program fought the status quo and forced real change: more first-time applicants showed up. |
| Mastodon | Officials said the new housing plan will be reviewed by parliament next month. | Parliament will review a new housing plan announced by officials next month—it's infuriating that our homes and futures are being bargained over. This is a moral decision about fairness and the future of our communities. |

***Table 2**: Illustrating examples of the rewrites produced by attention optimization.*

### ***Model-Based Attention Optimization Selects Different Political Forms***

When the estimated price function becomes the criterion for selecting among generated candidates, semantic constraint has a clear cost. Strict-factual candidates increase predicted reward by .087 on X/Twitter, .095 on Bluesky, and .098 on Mastodon, while preserving mean cosine similarity between .937 and .942. Attention-seeking selection increases reward by .396, .457, and .591 respectively, while similarity falls to .715–.736. The less constrained regime therefore produces approximately four to six times the reward gain within the model-based selection task.

Figure 3 places this reward gain and semantic loss on the same trajectory and then decomposes the accompanying style change. Relative to strict-factual candidates, attention-seeking rewrites increase personalization on all three platforms, moral-emotional framing most strongly on Mastodon, social positioning on all platforms, and incivility most strongly on Bluesky and X/Twitter. Neutral descriptions become situated performances: a speaker appears, a collective is invoked, an adversary is named, or an administrative issue becomes a moral conflict. These rewrite-style indicators should not be read as identical to the observational feature families; they describe how the generated candidates changed under looser constraints.

The result should not be directly interpreted as evidence that platforms cause misinformation, as the generated candidates were not fact-checked. What the probe suggests is that greater semantic and framing drift makes larger gains available. This suggests an incentive environment in which factual discipline and attention optimization can point in different directions.

### ***Robustness***

The main robustness checks are reported in the Supplemental Materials. Reward-model accuracy remains positive in every follower-count quartile for all nine platform-outcome models. On X/Twitter, performance among authors absent from training is close to full-test performance across outcomes (Pearson $r$ = .292-.366). Author-held-out TF-IDF models also retain signal across platforms, whereas within-author label permutations and shuffled-label controls are approximately null. These checks do not eliminate community or topic confounding, but they reject a purely memorized author-style explanation.

## Discussion

Most empirical research on social media and politics asks what platforms deliver to users and what this exposure does to their attitudes, polarization, and behavior. This paper argued for shifting the attention upstream. It has proposed an exploratory method to study mediatization through the attention price signal: the structured returns through which platformed communication environments make some forms of expression more advantageous than others. Our findings suggest that this signal is measurable: text predicts engagement even after adjusting for major audience and timing advantages, and approval and amplification emerge as distinct, sometimes divergent, reward systems.

Approval is the more portable market. Across platforms, likes reward personal stories, affinity, respect, compassion, and positive emotion, consistent with a general social logic of endorsement. Amplification is more contingent and more politically consequential. X/Twitter sharing penalizes relational, positive, and epistemically qualified language while rewarding insult, identity attack, and toxicity. Bluesky reposts favor neutral, lower-emotion, less personal language. Mastodon boosts reward reasoning, nuance, compassion, and collective positioning, while also showing smaller positive returns to insult and identity attack. X impressions form a further profile, closer to analytical reach than to reposting. These differences suggest that engagement cannot be treated as a single outcome: approval and amplification price different political acts.

The production perspective shifts attention from polarized audiences to political suppliers. Elites structure conflict through frames, cues, and identities (Levendusky, 2009; McCarty et al., 2006). Attention markets add an incentive account of how those structures are reproduced and intensified. Where identity attack or emotionally legible conflict earns circulation, actors may receive a recurrent signal that makes intergroup antagonism appear communicatively advantageous. This does not require a deliberate decision to polarize. Producers can copy successful formats, retain formulations that performed well, and internalize even hostile engagement as success (Mears & Beauvais, 2025). The mechanism is general, but its direction is not uniform. It depends on the platform, the audience, and the engagement action through which attention is allocated.

X sharing most clearly prices the communicative grammar associated with polarizing and populist discourse. Identity attack and insult receive positive returns, while respect, affinity, reasoning, and nuance receive negative returns (Finkel et al., 2020; Iyengar et al., 2019). Alternative and less engagement-optimizing platforms display different reward structures. Mastodon's positive returns to collective language, reasoning, nuance, and compassion do not fit a simple polarization account, and Bluesky reposts provide little linear evidence of an incivility premium. Yet this does not mean that incivility disappears outside X. Nonlinear models suggest that toxicity is rewarded for X/Twitter and Bluesky likes up to observed peaks near mean toxicity .14 and .31, respectively, before returns decline. These peaks sit above routine discourse but below the most toxic tail. Mastodon, by contrast, has an elevated but sparse high-toxicity tail rather than a clearly identified interior optimum. The evidence therefore supports a bounded and outcome-specific incivility premium: incivility is rewarded most clearly in X/Twitter sharing and in nonlinear approval curves for X/Twitter and Bluesky likes, but not as a universal rule that more abuse always produces more attention.

The computational rewriting probe makes one possible path of adaptation visible. When predicted reward becomes the selection criterion, larger gains become available as fidelity constraints loosen. Selected messages acquire clearer speakers, constituencies, moral stakes, and adversaries. This does not show that human producers would choose the same revisions, nor that platforms cause misinformation; the rewrites were not independently fact-checked. It suggests something narrower but theoretically important: close preservation of propositional content limits predicted attention gains, while greater semantic and framing latitude permits larger gains by allowing messages to move toward personalization, moral-emotional framing, social positioning, and conflict. In platformed attention markets, fidelity can become costly.

Several limitations should be emphasized. Reward models estimate associations, not causal effects of wording. Topic, community, reputation, unobserved exposure, user cultures, and platform-specific sampling processes may remain entangled with textual features. Platform samples differ in period and population, limiting direct comparison; Mastodon engagement may be incomplete across federation boundaries; and the scoring models may encode measurement bias. The analysis also cannot decompose the observed price signal into separate contributions from algorithmic ranking, affordances, audience heuristics, status hierarchies, and exogenous events. Most importantly, the analysis measures price signals rather than actual producer adaptation. It identifies the structured incentive environment in which adaptation can occur, not the extent to which particular politicians, journalists, or organizations respond to it. Panel studies of producers, fixed-exposure message experiments, institutional research, and ethnographies of political content production are needed to establish who learns these signals, how they interpret them, and how strongly they adapt.

## Conclusion

The politics of social media cannot be fully understood by tracing what platforms deliver to audiences, because platformed environments also shape the conditions under which politics is produced. By rendering attention visible, comparable, and consequential, they create markets in which expression receives differentiated returns, and these returns provide recurring signals to actors for whom visibility is a key condition of political relevance. Because those signals appear to come from "the public," they can be read not merely as platform feedback, but as evidence of what people notice, value, oppose, or demand.

The findings suggest that the less engagement-optimized platforms Mastodon and Bluesky reward more constructive forms of politics than X/Twitter: Mastodon rewards reasoning, nuance, compassion, and collective positioning, while Bluesky reposts show little evidence of a straightforward incivility premium. Yet neither platform is free from contentious or toxic incentives; Mastodon contains a sparse high-toxicity tail, and Bluesky likes reward toxicity up to a bounded optimum. While alternatives to X appear to encourage more reasoned and nuanced discourse, removing engagement algorithms is not a cure for uncivil and conflictual discourse.

The diverging reward signal across platforms also points to a possible future shift in the relationship between social media and politics, as publics increasingly sort across platforms along partisan lines (Törnberg, 2026b). If politicians come to depend on different platform publics, they may encounter different reward pressures shaping their incentives and their reading of the electorate: one environment may favor antagonistic identity performance, another may reward institutional critique, collective language, or deliberative style. Platform fragmentation could therefore produce not one unified media logic, but a set of partisan and subcultural attention markets that pull political actors in different directions.

The paper has suggested an empirical approach to studying the mediatization of politics that shifts attention from how citizens are affected by social media to what kinds of politics platformed attention environments make advantageous for political suppliers. This also shifts the imagination of platform intervention. Distributional approaches suggest primarily consumer-protection approaches: shielding citizens from harmful content or improving what algorithms place before them. A mediatization perspective points to a complementary political-economic question: how attention is allocated, how metrics and audiences make value visible, and what kinds of politics the resulting environment makes advantageous. The central question is thus how the attention economy can be organized so that deliberative, substantive, and civically ambitious politics is not merely possible, but rewarded.